\documentclass[prl,superscriptaddress,floatfix,nofootinbib,twocolumn,aps,10pt]{revtex4-2}
\usepackage{graphicx,amssymb,amsmath,bm,dcolumn,xcolor,mathtools}

\usepackage{upgreek}
\usepackage{comment}
\usepackage{xfrac}
\usepackage{appendix}
\usepackage{enumitem}
\usepackage[normalem]{ulem}
\usepackage{soul}
\newcounter{parnum}
\usepackage[colorlinks,linkcolor=blue,citecolor=blue,urlcolor=blue]{hyperref}

\newif\ifshowedits
\showeditstrue   

\begin{document}

\title{Dual mass milligram-scale torsion oscillator for vibration-free optomechanical sensing}

\author{J. Manley}
\affiliation{National Institute of Standards and Technology, Gaithersburg, MD 20899, USA}

\author{T. Bsaibes}
\affiliation{National Institute of Standards and Technology, Gaithersburg, MD 20899, USA}
\affiliation{Department of Physics, University of Maryland, College Park}

\author{C. A. Condos}
\affiliation{Wyant College of Optical Sciences, University of Arizona, Tucson, AZ 85721, USA}

\author{W. A. Terrano}
\affiliation{Department of Physics, Arizona State University, Tempe, AZ 85281, USA}

\author{D. J. Wilson}
\affiliation{Wyant College of Optical Sciences, University of Arizona, Tucson, AZ 85721, USA}

\author{J. R. Pratt}
\affiliation{National Institute of Standards and Technology, Gaithersburg, MD 20899, USA}

\begin{abstract}
Chip-scale optomechanical devices are driving the miniaturization of inertial sensors and next generation fundamental physics experiments. However, precision at the theoretical limit is often unattainable due to extraneous vibrations. One solution is tailoring the device to isolate a degree of freedom from the environment while maintaining coupling to the signal of interest. To this end, we introduce a dual milligram-mass torsion oscillator, formed by mass loading a strained silicon nitride nanoribbon. The antisymmetric torsion mode suppresses vibrations by over an order of magnitude to achieve a thermally limited torque sensitivity of $10^{-18}$~Nm/$\sqrt{\rm Hz}$ while maintaining ultralow loss. We demonstrate the sensing ability by detecting an optical radiation pressure torque of $10^{-16}$~Nm over a 30 Hz bandwidth. We also characterize the device for frequency-based gravimetry, demonstrating $10^{-6}g_0$ ($g_0=9.8$~$\rm m\,s^{-2}$) precision in 30 seconds with an oscillation amplitude of only 100~$\upmu$rad. This device demonstrates a technique for overcoming vibration noise, with broad implications for optomechanical sensing from commercial applications to fundamental physics experiments.
\end{abstract}

\maketitle

\section{Introduction}

\begin{figure*}[!t]
    \centering
    \includegraphics[width=\textwidth,trim= 0in 0in 0in 0in]{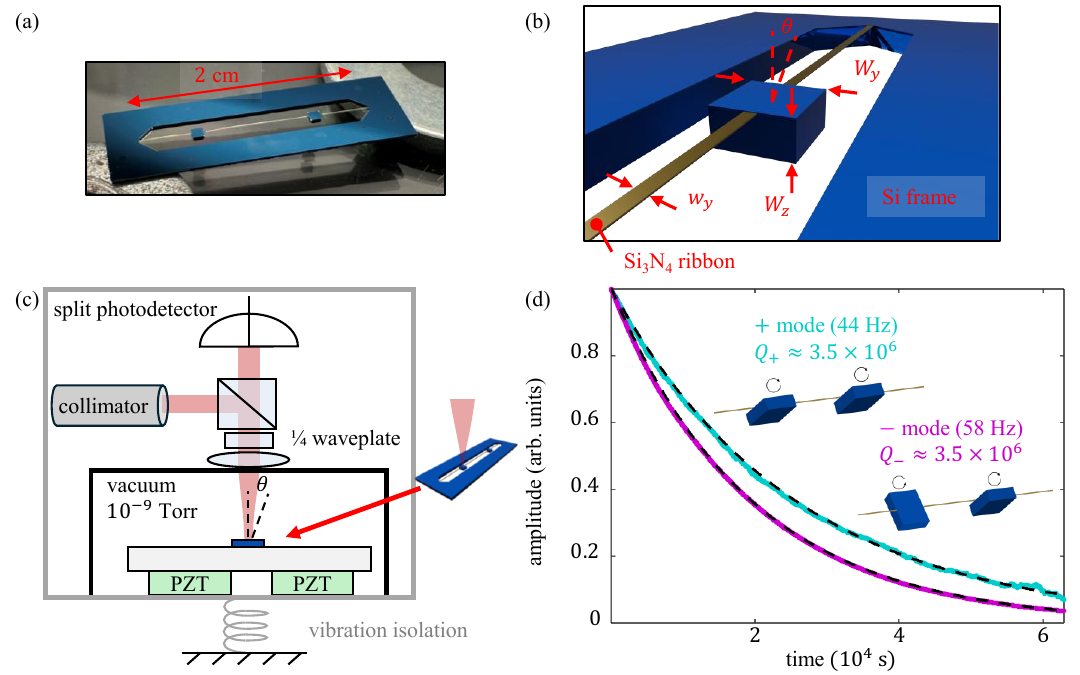}
    \caption{Dual test mass torsion oscillator. (a) Photograph of a fabricated device. (b) Rendering of the device, zoomed in on a single test mass. (c) Illustration of the experimental apparatus. PZT: Piezoelectric transducer. (d) Ringdown measurements of quality factors, performed non-simultaneously. The dashed black lines are fits proportional to $e^{-\omega_\pm t/2 Q_\pm}$.}
    \label{fig_device}
\end{figure*}

\begin{figure*}[!t]
    \centering
    \includegraphics[width=\textwidth,trim= 0in 0in 0in 0in]{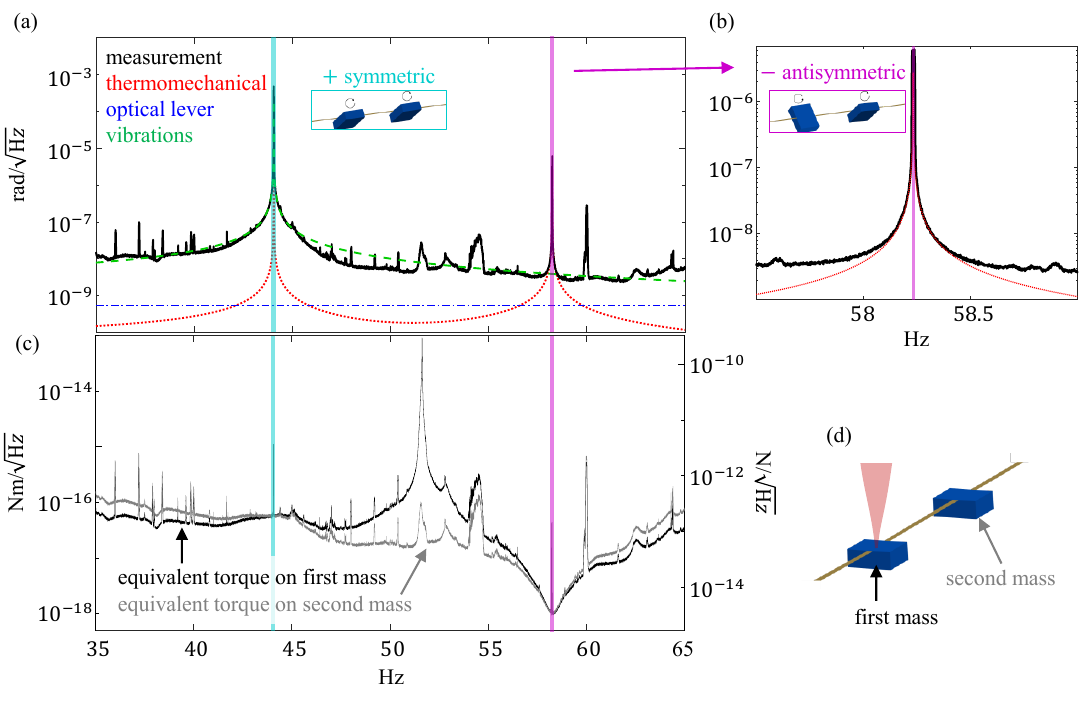}
    \caption{(a) Measured free running angular displacement spectrum. Dashed curves are models for thermomechanical noise (red), laboratory vibrations coupling to the symmetric mode (green), and optical measurement noise (blue). (b) Displacement spectrum near antisymmetric mode resonance. (c) Equivalent torque noise, considering a torque signal applied to either test mass. The right axis is the equivalent force noise assuming a local point force applied with a lever arm of $0.3$ mm. (d) Rendering of device, defining the masses mentioned in (c).}
    \label{fig_PSD}
\end{figure*}

\begin{figure*}
    \centering
    \includegraphics[width=2\columnwidth,height=0.77\textheight,keepaspectratio,clip,trim= 0in 0in 0in 0in]{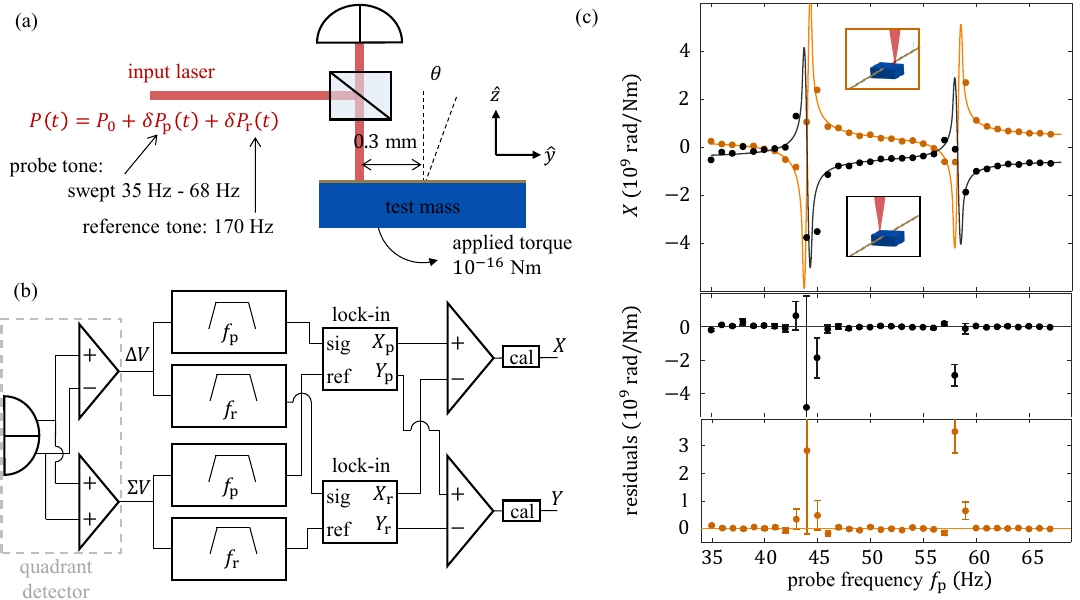}
    \caption{Torque sensitivity characterization with optical radiation pressure. (a) Illustration of the experiment. An intensity-modulated laser both applies an oscillating torque and measures the angular response. (b) Diagram of digital postprocessing. The sum and difference signals from the photodetector are used for lock-in detection of the oscillator response to the probe tone $\delta P_\text{p}$, using a second tone $\delta P_\text{r}$ for common noise cancellation. The tones are distinguished with bandpass filters: 33 Hz to 70 Hz for the swept probe; 167.5 Hz to 172.5 Hz for the constant reference. (c) Top: Frequency response function. Black (orange) points indicate the measured response when the laser beam is positioned approximately 0.3 mm ($-0.3$ mm) from the torsion axis. Solid lines are model fits. Middle and bottom plots display fit residuals and error bars indicate the standard error. }
    \label{fig_rp}
\end{figure*}

\begin{figure*}
    \centering    \includegraphics[width=2\columnwidth,height=0.77\textheight,keepaspectratio,clip,trim= 0in 0in 0in 0in]{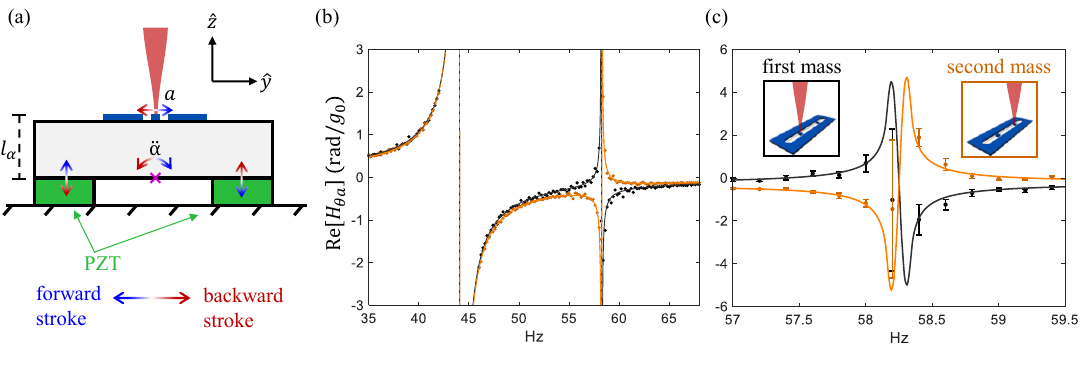}
    \caption{Vibration sensitivity characterization through driven response. (a) Illustration of the experiment, where two piezoelectric transducers driven out of phase tilt the sample holder, applying a horizontal acceleration to the device. The experiment is repeated after translating the optics from one test mass to the other. (b) In-phase response of the device. The model fits imply a 13-fold lower sensitivity of the antisymmetric mode. (c) Response near antisymmetric mode resonance.}
    \label{fig_accel}
\end{figure*}

\begin{figure}
    \centering
    \includegraphics[width=1\columnwidth,trim= 0in 0in 0in 0in]{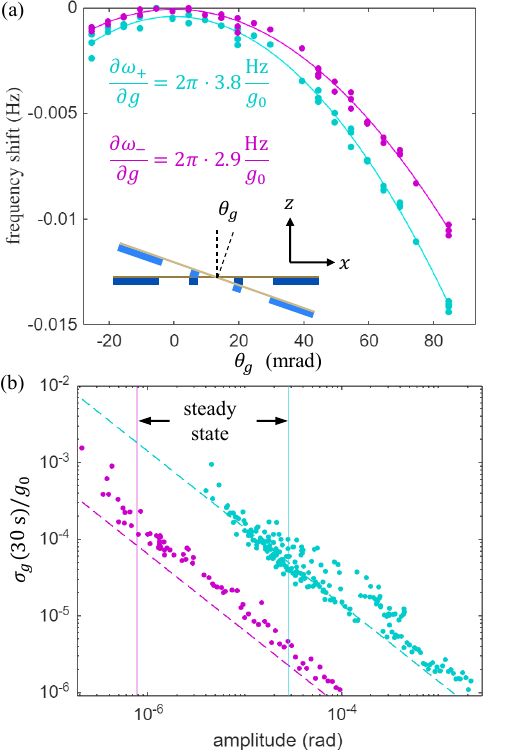}
    \caption{Gravity sensitivity characterization. (a) The parametric frequency sensitivity to gravity measured by tilting the experiment platform to simulate a change in gravity. Inset: Illustration of the direction of rotation of the device. (b) The gravity resolution of each mode, for a 30 s long measurement time, measured as a function of oscillation amplitude after initial excitation with the PZTs. Vertical lines mark the mean steady state amplitude for unexcited oscillators. }
    \label{fig_gravity}
\end{figure}

Micromechanical sensors at the milligram scale are an active area of research for gravity measurement. Motivated by compactness, mobility, and low cost, several devices have measured Earth's tides~\cite{middlemiss2016measurement,prasad202219,mustafazade2020vibrating,wang2025measurement,gao2026force,tang2019high}, with more showing promise for higher frequency inertial sensing~\cite{jiao2024optomechanical,li20222,weidong2025low,zhou2021broadband,guzman2014high,bawden2025precision}. 

Fundamental gravitational science has likewise seen a push toward the milligram scale. A prominent driver is measuring gravity at short distance, where experiments below 100~$\upmu$m are motivated for physics beyond the Standard Model~\cite{adelberger2003tests,adelberger2009torsion,manley2024microscale}, such as fifth forces~\cite{kaplan2000couplings,antoniadis2003brane,arkani1999phenomenology} and theories for dark energy~\cite{Kapner2007,sundrum2004fat,adelberger2007implications,upadhye2012chameleon,montero2023dark}. While experiments have been performed with a variety of sensing platforms, including micromechanical~\cite{long2003upper,chen2016stronger,geraci2008improved,fischbach2001new} and levitated sensors~\cite{blakemore2021search,venugopalan2026optomechanical}, so far only traditional torsion balances have succeeded in measuring the gravitational interaction within $100$ $\upmu$m~\cite{Kapner2007, lee2020new}. The sizes of sensors used in leading experiments are distinct: at a length scale of 7~$\upmu$m, constraints from $\upmu$g-scale test masses used to probe shorter distance~\cite{chen2016stronger} meet constraints from gram-scale torsion balances~\cite{lee2020new,tan2020improvement} that dominate at longer range, suggesting mg-scale experiments for bridging the gap. Further motivation comes from a surge in the development of novel fabrication techniques~\cite{catano2020high,bsaibes2025lithographically,manley2026nanofabricated,cong2021chip} to help accelerate the miniaturization of classical torsion balances~\cite{agafonova2026one,westphal2021measurement,komori2020attonewton,sokhi2024all,guan2026all}.

Mass-loaded silicon nitride (Si$_3$N$_4$) nanoribbon torsion oscillators~\cite{pratt2023nanoscale} have emerged as promising platforms for optomechanical inertial sensing. By exploiting strain- and gravity-induced dissipation dilution, their torsional modes can achieve ultralow loss. Optical lithography enables precise control over the nanoribbon and suspended mass dimensions. Cavity-free displacement readout is provided by an optical lever, and the device motion is controllable with radiation pressure. These advantages enable potential sensing applications from dark matter searches~\cite{dey2026optomechanical} to compact frequency-based gravimetry~\cite{condos2025ultralow}. The devices are particularly well suited for short range gravity experiments~\cite{manley2024microscale}, where the optically flat test masses enable close approach, consolidating a large proportion of the test mass to within the interaction range.

A central challenge in optomechanical gravity experiments is vibration noise, where incoherent excitation from anthropogenic or seismic activity can overshadow weak signals. In tests of short range gravity, the ability of cm-scale torsion balances to achieve shorter surface separations is limited by vibration-induced electrostatic noise~\cite{lee2020new,tan2020improvement}. More compact MEMS~\cite{geraci2008improved,smullin2005constraints} and levitated~\cite{blakemore2021search} sensors also experience sensitivity limitations from ambient vibrations and direct vibrational coupling to the modulated source mass. These environmental disturbances are a pervasive issue beyond fundamental science, affecting both atom-interferometer absolute gravimeters~\cite{wu2024construction,zhou2025stability,des2025gravity} and MEMs-based relative gravimeters~\cite{middlemiss2016measurement,prasad202219,mustafazade2020vibrating}. While background vibrations can be mitigated with a vibration isolation system~\cite{matichard2015seismic} or by working in a low noise laboratory (such as underground~\cite{farah2014underground} or in a cave~\cite{wu2024construction,gao2026force,tang2019high}), these solutions are not viable for compact, field-deployable gravimeters.

Vibrations from the environment have been shown to dominate the noise floor in studies of Si$_3$N$_4$-based optomechanical torsion oscillators~\cite{condos2025ultralow,manley2024microscale}. Unlike traditional torsion balances, a key feature of these oscillators is that the center of mass is offset from the torsion axis, allowing them to be used for frequency-based gravimetry~\cite{condos2025ultralow}, but also making them sensitive to lateral acceleration noise~\cite{manley2024microscale,chandak2026nano}. Using feedback actuation to drive large coherent motion can improve the signal-to-noise ratio against the background vibrations, at the expense of introducing nonlinear effects~\cite{condos2025ultralow,pratt2023intersection}. For a short range gravity experiment, one could engineer a new fabrication process to enable symmetric mass-loading about the torsion axis, albeit while forfeiting gravitational dissipation dilution. 

In this work we introduce a method for mitigating vibration noise in optomechanical torsion oscillators with the inclusion of two test masses on the same suspension. Similar double-paddle oscillators have been explored to achieve low loss in silicon devices, and it has been observed that their antisymmetric torsion modes are less sensitive to external vibrations~\cite{liu2001modes}. Here, the coupled masses experience an antisymmetric oscillation mode with no translational motion of the center of mass, decoupling from translational accelerations of the device frame. As a result, the 58 Hz antisymmetric oscillation mode demonstrates precision at the thermal noise limit, paving the way for deployment in a variety of sensing applications. The following text describes a series of experiments to characterize the device, its noise floor, susceptibility to vibration noise, and sensitivity to applied torques and relative shifts in gravity, and the Appendices provide experimental details and models of the device mechanics.

\section{Results} \label{sec:results}
The device is fabricated from a 400-$\upmu$m-thick silicon wafer, double-side-coated with 100 nm of high-stress (roughly 1 GPa) Si$_3$N$_4$, photographed in Fig.~\ref{fig_device}a and sketched in Fig.~\ref{fig_device}b.  The $m=0.93$ mg Si test masses remain after etching the surrounding substrate to release the Si$_3$N$_4$ ribbon. The dual test mass system has two modes of torsional oscillation, defined by the symmetric ($+$) and antisymmetric ($-$) angular rotation of the test masses about the ribbon central axis. The resonance frequencies are $\omega_+  = 2\pi \cdot 44$ Hz and $\omega_-=2\pi \cdot 58$ Hz, with restoring torques provided by both ribbon stress and Earth's gravity acting on the suspended masses~\cite{pratt2023nanoscale}.

Figure~\ref{fig_device}c depicts the experimental setup. Within a vacuum chamber, the chip sits atop a holder that can be actuated using piezoelectric transducers for feedback cooling and driving the resonator. The angular motion of the test masses is measured using an optical lever, where a 1.55~$\upmu$m wavelength Gaussian laser beam is focused with a 5 cm focal length lens onto one of the test masses, and the deflection of the reflected light is measured with a quadrant photodetector. The laser spot is typically focused near the center of the test mass, as opposed to near the edge as in the experiment described in Section~\ref{sec:rp} (Fig.~\ref{fig_rp}) where a radiation pressure torque is desired.

\subsection{Quality factor measurement}
Low mechanical loss is a key advantage of the device, as dissipation is responsible for the thermomechanical noise that sets the ultimate limit on sensitivity~\cite{saulson1990thermal}. Ringdown measurements of the mechanical quality factors $Q_\pm$ are displayed in Fig.~\ref{fig_device}d, where the free exponential decay in oscillation amplitude is observed after initial excitation. Each ringdown was performed separately, where in each case piezoelectric transducers were used to simultaneously excite one mode to a large amplitude and damp the other in order to isolate a single mode and prevent cross-contamination of the signals. For the measurement of the 44 Hz mode depicted by the cyan curve, the 44~Hz (58~Hz) mode was initially excited (damped) to about 2 mrad (6~$\upmu$rad). For the measurement of the 58~Hz mode depicted by the magenta curve, the 58~Hz (44~Hz) mode was initially excited (damped) to about 4 mrad (80~$\upmu$rad). For both modes, $Q_\pm \approx 3.5 \times 10^6$ was observed.

\subsection{Noise characterization}
The limit on sensor precision comes from various noise sources, including readout noise in the optical lever and incoherent motion of the test masses driven by thermomechanical noise and laboratory vibrations.

Figure~\ref{fig_PSD}a shows the power spectral density (PSD) of the optical lever measurement of the angular displacement $\theta$, where the torsion modes are visible at 44 Hz and 58 Hz. Optical readout noise does not appear to be significant in the 35 Hz to 70 Hz band; readout noise contributes roughly $5\times 10^{-10}$ $\rm rad/\sqrt{\rm Hz}$ (blue dashed curve), which we have estimated with a separate reference measurement in which the probe beam was reflected off the stationary frame of the device chip. 

The symmetric motion is predominantly driven by extraneous laboratory vibrations, which may arise from both linear and angular displacement of the device platform (see Appendix~\ref{app_sec:vibrations}). The vibration spectrum is not white and contains various spurious peaks, but can be roughly approximated by linear accelerations at the level $1.5\times 10^{-6}$~$g_0/\sqrt{\rm Hz}$ or tilts at the level $3.3$~nrad/$\sqrt{\rm Hz}$ exciting the symmetric mode to a root mean square oscillation amplitude of about $40$ $\upmu$rad (green dashed curve). 

Thermomechanical torque noise sets the fundamental noise floor. To estimate thermomechanical noise, we adopt a structural damping model~\cite{saulson1990thermal} for each mode, giving rise to torque noise $S_{\tau_\pm}^\text{th}=4 k_\text{B} T I \omega_\pm^2/\left(\omega Q_\pm \right)$ through the fluctuation dissipation theorem~\cite{saulson1990thermal}, where $k_\text{B}$is the Boltzmann constant and $T$ is temperature. This translates to an angular displacement PSD $S_\theta(\omega) = \left| \chi_+(\omega)\right|^2 S_{\tau_+}^\text{th}(\omega)+\left| \chi_-(\omega)\right|^2 S_{\tau_-}^\text{th}(\omega)$ where 
\begin{equation} \label{eq:susceptibility}
    \chi_\pm(\omega)= \frac{I^{-1}}{\omega_\pm^2 - \omega^2 - i \omega_\pm^2 /Q_\pm}
\end{equation}
is the mechanical susceptibility and $I$ is the moment of inertia of each test mass about the ribbon central axis (see Appendix~\ref{app_sec:mechanics}). In the antisymmetric mode, we observe a mean oscillation amplitude of about $0.8$~$\upmu$rad, in reasonable agreement with the a priori predicted thermal amplitude $\sqrt{2 k_\text{B} T / (I \omega_-^2)}=0.7$~$\upmu$rad at room temperature, as can be seen by zooming in on the PSD around 58~Hz in Fig.~\ref{fig_PSD}b.


The equivalent torque noise spectrum is displayed in Fig.~\ref{fig_PSD}c, transformed from the displacement spectrum in Fig.~\ref{fig_PSD}a using
\begin{equation}\label{eq:H_tau}
    H_{\theta\tau} =
    \begin{cases}
        \dfrac{\chi_+ + \chi_-}{2}, & \text{torque on mass 1} \\[6pt]
        \dfrac{\chi_+ - \chi_-}{2}, & \text{torque on mass 2}
    \end{cases}
\end{equation}
which describes the susceptibility to a torque applied locally to each of the test masses, $S_\theta = \left|H_{\theta \tau} \right|^2 S_\tau$. The torque noise sensitivity at the 58~Hz antisymmetric resonance is at the thermal limit, $10^{-18}$~Nm/$\sqrt{\rm Hz}$. A key feature in Fig.~\ref{fig_PSD}c is the peak near 52 Hz, where the measurement is insensitive to an applied torque on the observed test mass (mass 1, depicted in Fig.~\ref{fig_PSD}d) due to destructive interference between the responses of the two modes. However, this extinction would not occur if the torque signal is applied to the other mass (mass 2). The sensitivity of the device as a force sensor is approximated using the right axis of Fig.~\ref{fig_PSD}c---scaled as an equivalent force applied at a lever arm of 0.3 mm from the torsion axis---reaching a value of $3$~fN/$\sqrt{\rm Hz}$ at the antisymmetric mode resonance.

\subsection{Torque sensitivity characterization with radiation pressure} \label{sec:rp}
Radiation pressure provides a convenient way to calibrate an optomechanical torque sensor~\cite{sokhi2024all,guan2026all}. We measured the frequency response of the device (model in Eq.~\ref{eq:H_tau}) to an oscillating radiation pressure torque applied by the detection laser, displayed in Fig.~\ref{fig_rp}. The laser spot is positioned a distance $l\approx 0.3$ mm from the torsion axis. The diode current is modulated to produce a fractional laser power modulation of $\xi_\text{p}\approx 10~\%$ at probe frequency $f_\text{p}$ (the 170 Hz reference modulation was about $2~\%$), thereby exerting an oscillating radiation pressure torque with amplitude $\tau=2 \xi_\text{p} P_R l/c\approx10^{-16}$ Nm, where $P_R\approx 0.6$ mW is the mean reflected laser power and $c=3\times 10^8$~m/s is the speed of light. The mechanical response is characterized by sweeping the probe frequency from 35 Hz to 68 Hz. 

The mechanical response was extracted from the optical lever signal using lock-in detection with the quadrant detector sum signal as a reference, as described in Fig.~\ref{fig_rp}b. To avoid cross-talk, we actively subtract a reference tone generated by injecting an auxiliary modulation tone at 170 Hz. The same procedure is performed for both the probe and reference tones, which are distinguished with bandpass filtering. The estimated response is corrected by subtracting the reference from the probe measurements. The experiment, which was repeated with the laser spot at two different positions on opposite sides of the torsion axis, was performed continuously---lasting 27 hours (64 hours) for the black (orange) points---with the probe tone scanning from 35 Hz to 68 Hz every 500 seconds. The measurement record was split into 15 second segments and the measured response was sorted into 1 Hz frequency bins based on the average probe tone frequency $f_\text{p}$ during the segment. After reference subtraction, each bin's measurements were averaged to produce the points in Fig.~\ref{fig_rp}c. The error bars indicate the standard error of measurements in each frequency bin. 

The result in Fig.~\ref{fig_rp}c demonstrates agreement with the model, where the measured in-phase response $X$ is compared to the real part of the predicted susceptibility in Eq.~\ref{eq:H_tau}. Faithful measurement on resonance is precluded by the narrow linewidths of the mechanical modes, as the frequency scan rate was too fast and slow drift of the resonance frequencies distorts the response. Off-resonance, the response function is predominantly real, and the model is fit to the real part of the measured response as $X\propto \text{Re}\left[H_{\theta\tau}\right]$ with error of less than about 30~\% (see Appendix~\ref{app_sec:rp_LI}).

\subsection{Vibration sensitivity characterization}\label{sec:vib_char}
In principle, the dual-mass design enables torque sensing with immunity to external vibrations. We characterized the vibration sensitivity with a driven response measurement, using PZTs to shake the sample. The principle of the experiment is sketched in Fig.~\ref{fig_accel}a, where an oscillating tilt $\alpha$ is applied to the sample holder by driving two PZTs out of phase. The tilt amplitude was approximately 45 nrad, confirmed by a separate measurement with the optical lever focused on the chip's frame.  Located at a lever arm of $l_\alpha\approx 1$ cm, the chip experiences a frequency-dependent acceleration $a=l_\alpha \omega^2 \alpha$ ranging from $10^{-6}$ $g_0$ to $10^{-5}$ $g_0$. 

The susceptibility of the dual-mass oscillator to external vibrations was characterized by scanning the PZT drive frequency between 35 Hz and 68 Hz and recording the angular response, as shown in Fig.~\ref{fig_accel}b-c. The band from 42~Hz to 46~Hz was intentionally avoided to prevent excitation of the 44 Hz symmetric mode beyond the dynamic range of the optical lever. The susceptibility of the angular displacement to purely translational acceleration along the horizontal ($y$) axis can be shown to be (see Section \ref{app_sec:vibrations})
\begin{equation}\label{eq:H_vib}
    H_{\theta a} =
    \begin{cases}
        \frac{m W_z}{2} \left( \eta_+ \chi_+ + \eta_- \chi_- \right), & \text{observing mass 1} \\[6pt]
        \frac{m W_z}{2} \left( \eta_+ \chi_+ - \eta_- \chi_- \right), & \text{observing mass 2}
    \end{cases}
\end{equation}
where the coupling to each mode is described by $\eta_\pm$. Note that while the device is sensitive to tilts in addition to horizontal acceleration, the response to tilt is an order of magnitude smaller in the experiment in Fig.~\ref{fig_accel} (see Appendix~\ref{app_sec:tilts} for the tilt sensitivity model). Ignoring fabrication imperfections, we expect $\eta_+= 1$ and $\eta_-=0$. Fits to the data demonstrate $\eta_+=1.06$ and $\eta_-=0.08$, suggesting the antisymmetric mode is 13 times less sensitive to translational vibrations. The non-zero sensitivity of the antisymmetric mode could arise due to uneven etching producing non-identical test masses differing in mass by $8~\%$. An alternative explanation is that the excitation contains a component of twist about the $z$-axis, which would drive antisymmetric motion without any fabrication imperfection (see Appendix~\ref{app_sec:twists}).

\subsection{Gravity sensitivity characterization}
The pendulum nature of the suspended masses enables parametric sensitivity to the local gravitational acceleration $g$, where relative fluctuations in $g$ shift the resonance frequencies (see Eq.~\ref{eq:dwdg} and Refs.~\cite{pratt2023nanoscale,condos2025ultralow}). In Fig.~\ref{fig_gravity}a the parametric sensitivity to gravity is measured by tilting the device platform an angle $\theta_g$ as indicated in the inset. The tilt changes the effective gravity as $g=g_0 \cos \theta_g$, and the change in frequency is measured. 

The antisymmetric mode achieves better gravity resolution than the symmetric mode owing to its reduced noise. On short timescales---less than about $10^2$ s---the resolution is limited by the ability to precisely measure the mode frequencies in the presence of incoherent oscillation driven by vibrations or thermomechanical torque noise. Given a measurement time $t_\text{ave}$, the resolution of a given mode is limited to~\cite{condos2025ultralow}
\begin{equation}
    \sigma_{g,\pm} = \left|\frac{\partial \omega_\pm}{\partial g}\right|^{-1} \sqrt{\frac{\omega_\pm}{2 Q_\pm t_\text{ave} \epsilon_\pm}} 
\end{equation}
where $\epsilon_\pm$ is the ratio of the oscillating power to the background noise (see Section \ref{app_sec:gravity}), such that better performance can be achieved by either driving large coherent oscillations or reducing the background noise. Both methods are illustrated in Fig.~\ref{fig_gravity}b, where the gravity resolution is calculated from the Allan deviation of $t_\text{ave}=30$~s long resonance frequency measurements within 1500 s time intervals for both modes during their own respective ringdowns and plotted against the average oscillation amplitude during that interval. Both modes demonstrated improved performance with oscillation amplitude, but the antisymmetric mode outperforms the symmetric mode by an order of magnitude due to the lower noise floor.

\section{Conclusion}
We have fabricated an ultralow loss, milligram-scale torsion oscillator that demonstrates reduced sensitivity to noise from laboratory vibrations. While both mechanical modes have exceptionally low damping rates and seismic-band resonant frequencies, the favorable geometry of the antisymmetric mode enables sensitivity at the thermomechanical torque noise limit. In comparison, the symmetric mode suffers a noise floor 40 times larger than thermomechanical noise. 

One of the main drivers for the development of mass-loaded Si$_3$N$_4$ torsion oscillators as torque sensors is to measure the gravitational interaction down to 25~$\upmu$m separation~\cite{manley2024microscale}. The technique introduced here to exploit the antisymmetric mode of the dual-mass system suppresses vibration noise, enabling sensing with a thermomechanically limited imprecision of $10^{-18}$ Nm/$\sqrt{\rm Hz}$ and the potential for improvement via cryogenic cooling. We demonstrated this sensing capability by applying and detecting an optical radiation pressure torque of $10^{-16}$~Nm. Looking forward, the torque sensing performance can be further improved by deploying a dual readout scheme, whereby measurement of the positions of both test masses can be combined to isolate the antisymmetric mode to increase the thermally limited bandwidth around resonance. 

Finally, we consider the potential performance as chip-scale relative gravimeters with parametric frequency sensitivity to gravity. 
A stable resonance frequency is crucial, and stochastic vibrations add noise to the frequency estimation. Previous attempts to overcome vibrations in the Si$_3$N$_4$-based platform by driving large amplitude motion have been hindered by the onset of mechanical and optical transduction nonlinearities~\cite{condos2025ultralow}. In this work, vibration noise has been reduced to access a larger signal-to-noise ratio, allowing the device to achieve $10^{-6} g_0$ precision in 30 seconds with an oscillation amplitude of only 0.1 mrad. This demonstrates an order of magnitude improvement in precision as a relative gravimeter by overcoming extraneous vibration noise via monitoring the antisymmetric mode.

\textit{Acknowledgments}---We are grateful to Atkin Hyatt and Mohanbabu Mani for providing the photograph in Fig.~\ref{fig_device}a, and to Daniel Barker for feedback on the manuscript. JPM acknowledges support from a NIST NRC Postdoctoral Research Associateship. DJW and CAC acknowledge support from NSF awards nos. 2239735 and 2330310, and from a UArizona National Labs Partnership Grant.

\appendix

\section{Mathematical Models}
\subsection{Torsion oscillation modes} \label{app_sec:mechanics}
The torsion constant of a stressed ribbon with length $l_\text{rib}$, width $w_y$, thickness $w_z$, and stress $\sigma$ is~\cite{pratt2023intersection}
\begin{equation} \label{eq:ksigma}
    \kappa_\sigma = \frac{w_z w_y^3}{12 l_\text{rib}}\sigma
\end{equation}
For the dual mass system, there is effectively a torsion spring of stiffness $\kappa_\sigma$ connecting each test mass to the frame and a torsion spring of stiffness $\kappa_\sigma/2$ connecting the two masses. The gravitational acceleration due to Earth $g$ provides an additional torsional stiffness on each mass $m$ of~\cite{pratt2023nanoscale}
\begin{equation}
    \kappa_g = \frac{1}{2} g m W_z
\end{equation}
where $W_z$ is the thickness of the masses. The dual mass system has two modes of oscillation. If $\theta_{i}$ is the angular displacement of test mass $i$, the symmetric (+) and antisymmetric (-) modes are defined by coordinates 
\begin{equation} \label{eq:phiModes}
    \phi_\pm\equiv \left(\theta_1 \pm \theta_2\right)/2
\end{equation}
with resonance frequencies
\begin{equation}
    \begin{aligned}
        \omega_+^2 &= \frac{\kappa_g + \kappa_\sigma}{I} \\
        \omega_-^2 &= \frac{\kappa_g +2 \kappa_\sigma}{I} 
    \end{aligned}
\end{equation}
where each test mass has mass $m=W_x W_y W_z \rho$ and moment of inertia $I=m\left(W_y^2/12 + W_z^2/3\right)$. We assume the density of silicon to be $\rho=2330$~kg/m$^3$.  

The thickness of the test masses and torsion ribbon are set by the original thickness of the Si wafer and Si$_3$N$_4$ film, with nominal values $W_z=400$ $\upmu$m and $w_z=100$ nm, respectively. The remaining dimensions are defined with optical lithography. The ribbon is designed to be $w_y=150$ $\upmu$m wide and span a gap of $w_x=2$ cm. The effective ribbon length in Eq.~\ref{eq:ksigma} is then $l_\text{rib}\approx w_x /4 = 5$ mm. The test masses are designed to be square with side length $W_x=W_y=1$ mm.

\subsection{Torque sensitivity}
In the frequency domain---with Fourier transform defined as $\tilde{x}(\omega)=\int dt\, e^{-i\omega t}x(t)$---the modal displacement due to modal torque is
\begin{equation}
    \tilde{\phi}_{\pm}(\omega)=\chi_\pm(\omega) \tilde{\tau}_\pm(\omega)
\end{equation}
where the mechanical susceptibility is
\begin{equation}
    \chi_\pm(\omega)= \frac{1}{I}\frac{1}{\omega_\pm^2 - \omega^2 - i \omega_\pm^2 /Q_\pm}
\end{equation}
assuming a structural damping model~\cite{saulson1990thermal}. Here, $Q_\pm$ is the quality factor. We often omit the Fourier frequency argument $(\omega)$ for brevity.

The modal torques can be understood as linear combinations of external torques applied to individual test masses, $\tau_\pm = (\tau_1 \pm \tau_2 )/2$. Additionally, thermomechanical torque noise affects the modes individually with PSD $S_{\tau_\pm}^\text{th}=4 k_\text{B} T I \omega_\pm^2/\left(\omega Q_\pm \right)$. We specifically target torque signals applied locally to just one of the test masses. For simplicity, we will always treat mass 1 as the receiver: $\tau=\tau_1$. The mode responses would be $\tilde{\phi}_{\pm}= \chi_{\pm} \tilde{\tau}/2$. 

The susceptibility to the signal torque, $H_{\theta \tau} = \tilde{\theta}/\tilde{\tau}$, varies depending on which mass is being measured. If measuring the first mass, $\theta=\theta_1$,
\begin{equation}
    H_{\theta\tau} = \frac{\chi_+ + \chi_-}{2} 
\end{equation}
If measuring the second mass, $\theta=\theta_2$, 
\begin{equation}
    H_{\theta\tau} = \frac{\chi_+ - \chi_-}{2} 
\end{equation}

\subsection{Sensitivity to vibrations} \label{app_sec:vibrations}
Seismic and anthropogenic activity drives an ever-present background of vibrations in the laboratory that may couple to the device through oscillating tilt and translation of the silicon frame. Here we present a model for how these motions affect the test mass motion.

\subsubsection{Translational (along $\hat{y}$) vibrations}
Linear accelerations perpendicular to the torsion axis (along the $y$-direction) $a$ effect a modal torque
\begin{equation} \label{eq:eta_def}
    \tau_\pm = \frac{m W_z}{2} \eta_\pm a
\end{equation}
where our design implies coupling coefficients $\eta_+=1$ and $\eta_-=0$, such that the antisymmetric mode is insensitive to such vibrations. However, in practice $\eta_-\neq 0$, as fabrication imperfections produce asymmetry in the test masses or ribbons. Generally, asymmetry in the device would be better described by appropriately defining the modes $\phi_\pm$ to ensure orthogonality by modifying Eq.~\ref{eq:phiModes} with unequal coefficients. For simplicity we will continue with the definition of Eq.~\ref{eq:phiModes} and account for the small asymmetry only in the sensitivity to vibrations as prescribed by Eqs.~\ref{eq:eta_def} and \ref{eq:zeta_def}. Translational vibrations appear in the measurement through susceptibility 
\begin{equation} \label{app_eq:theta_a}
    H_{\theta a}\equiv \tilde{\theta}/\tilde{a}=\frac{m W_z}{2} \left( \eta_+ \chi_+ + \eta_- \chi_- \right)
\end{equation}

\subsubsection{Tilting (about $\hat{x}$) vibrations} \label{app_sec:tilts}
Vibrations may also tilt the chip by angle $\alpha_x$ in Earth's reference frame, exerting a modal torque via the outer tethers:
\begin{equation} \label{eq:zeta_def}
    \tau_\pm = \zeta_\pm^x \kappa_\sigma \alpha_x
\end{equation}
where our design implies coupling coefficients $\zeta_+^x=1$ and $\zeta_-^x=0$. As for translational acceleration, asymmetry from fabrication imperfection may result in non-zero coupling to the antisymmetric mode, $\zeta_-^x\neq 0$. Rotational vibrations appear in the measurement through susceptibility 
\begin{equation} \label{app_eq:theta_tilt}
    H_{\theta \alpha_x}\equiv \tilde{\theta}/\tilde{\alpha}_x= \kappa_\sigma \left( \zeta_+^x \chi_+ + \zeta_-^x \chi_- \right)
\end{equation}
Generally, vibrations may tilt the whole experimental platform, including the optical lever. In this case, the susceptibility is $H_{\theta \alpha_x}= \kappa_\sigma \left( \zeta_+^x \chi_+ + \zeta_-^x \chi_- \right)-1$ when treating $\theta$ as the measured displacement rather than the true displacement in an inertial frame.

\subsubsection{Twisting (about $\hat{z}$) vibrations} \label{app_sec:twists}
While the antisymmetric mode is designed to be immune to translational vibrations---and tilt about $\hat{x}$---twists of the chip within the horizontal plane (about the $z$-axis) would excite to the antisymmetric motion. Given a twist angle $\alpha_z$ about the $z$-axis, centered between the test masses, each test mass would experience a torque with opposite sign and equal magnitude of roughly $m W_z l_\text{rib} \ddot{\alpha}_z /2$. Twisting vibrations would appear in the measurement through susceptibility
\begin{equation} \label{app_eq:theta_twist}
    H_{\theta \alpha_z}=\frac{m W_z l_\text{rib}}{2 } \omega^2\left( \zeta_+^z \chi_+ + \zeta_-^z \chi_- \right)
\end{equation}
where the model predicts coupling coefficients $\zeta_+^z=0$ and $\zeta_-^z=1$. 

It is experimentally challenging to introduce vibrations of a known polarization in order to isolate and fully characterize each susceptibility in Eqs.~\ref{app_eq:theta_a}, \ref{app_eq:theta_tilt}, and \ref{app_eq:theta_twist}. In the main text we simplify the analysis by assuming that the vibrations are purely translational (susceptibility in Eq.~\ref{app_eq:theta_a}), and treat the non-zero response of the antisymmetric mode as a fabrication imperfection resulting in $\eta_-\neq 0$. However, an alternative explanation is that the platform also experiences a twist $\alpha_z$, and that the non-zero response of the antisymmetric mode derives from $H_{\theta \alpha_z}$ with $\zeta_-^z=1$.  It can be shown that the twist needed to produce the measured response in Fig.~\ref{fig_accel} of the main text is 
\begin{equation}
    \alpha_z = \eta_- \frac{l_{\alpha_y}}{l_\text{rib}} \alpha_y = 0.08 \left(\frac{1{\rm \, cm}}{5 \rm \, mm}\right) \left(45{\,\rm nrad}\right) = 7\,\text{nrad}
\end{equation}
Unfortunately, the optical lever cannot measure $z$-axis rotations of the chip to determine whether the antisymmetric mode sensitivity to vibrations should be attributed to multiple vibration polarizations or simply fabrication imperfection.

\subsection{Parametric sensitivity to gravity} \label{app_sec:gravity}
The gravitational component of the torsion constant enables each mode to act as a clock gravimeter~\cite{condos2025ultralow} with parametric sensitivity
\begin{equation} \label{eq:dwdg}
    \begin{aligned}
        \frac{\partial \omega_+}{\partial g} &= \frac{\omega_+}{2g} \frac{\kappa_g}{\kappa_g + \kappa_\sigma} \\
        \frac{\partial \omega_-}{\partial g} &= \frac{\omega_-}{2g} \frac{\kappa_g}{\kappa_g + 2\kappa_\sigma}
    \end{aligned}
\end{equation}
The gravity resolution is limited by frequency uncertainty due to thermomechanical noise, which is 
\begin{equation}
    \sigma_{\omega_\pm} = \sqrt{\frac{\omega_\pm}{2Q_\pm \epsilon_\pm  t_\text{ave}}}
\end{equation} 
in measurement time $t_\text{ave}$~\cite{sadeghi2020frequency}. Here, $\epsilon_\pm \equiv \left< \phi_\pm^2\right>/\left<\phi_{\pm,\text{th}}^2 \right>$ is the ratio of the total mean square displacement to that of the background noise. The noise can be expressed as $\left<\phi_{\pm,\text{th}}^2 \right>=k_\text{B} T_{\pm,\text{eff}} / (I \omega_\pm^2)$, where $T_{\pm,\text{eff}}$ is the effective temperature of the mode, and it may exceed the physical temperature of the device if vibrations dominate over thermomechanical noise. The limit on the gravity resolution for each mode is
\begin{equation}
    \sigma_{g,\pm} = \left|\frac{\partial \omega_\pm }{\partial g} \right|^{-1} \sigma_{\omega_\pm}
\end{equation}

\section{Smoothing of susceptibility models near resonance}
The mechanical susceptibilities $\chi_\pm$ have a sharp feature near resonance due to the high quality factors of the modes. As a result, characterizing the functions $H_{\theta \tau}$ and $H_{\theta a}$ as described in Sections~\ref{sec:rp} and \ref{sec:vib_char} is difficult near resonance, as slow drift in the resonance frequencies during the measurement as well as the continuous scanning of the drive frequency effectively smear the resonance peak. We approximately account for this effect in the models plotted in Figs.~\ref{fig_rp} and \ref{fig_accel} by smoothing over a finite bandwidth $\Delta_\omega$ with a window function $w(\omega)$. For example, the smoothed model $H'(\omega)$ of a function $H(\omega)$ is
\begin{equation}
    H'(\omega) = \int_{\omega-\Delta_\omega/2}^{\omega+\Delta_\omega/2} d\Omega\, w(\Omega) H(\Omega)
\end{equation}
where the window is normalized as 
\begin{equation}
    \int_{\omega-\Delta_\omega/2}^{\omega+\Delta_\omega/2} d\Omega\, w(\Omega)=1
\end{equation}
A rectangular window is prone to edge effects yielding unrealistically sharp features. We use a symmetric Hann window, and note that the fits are largely unaffected by this choice due to the low number and high uncertainty of data points near resonance. We use a bandwidth of 1 Hz in Fig.~\ref{fig_rp} and 0.2 Hz in Fig.~\ref{fig_accel}, corresponding to the spacing of the frequency bins.

\section{Calibrating the optical lever} \label{sec:OL_calibration}
The optical lever depicted in Fig.~\ref{fig_device} uses a lens of focal length $L_\text{len}$ to focus the collimated laser beam onto the test mass. The reflected beam is deflected by an angle $2\theta$, producing a displacement $x_\text{d}=2 L_\text{len} \theta$ of the laser spot on the split photodetector. The difference voltage $\Delta V$ between the two halves of the photodetector ultimately provides an estimate of the angular displacement, where $\partial \Delta V / \partial \theta = 2 L_\text{len} \left(\partial \Delta V / \partial x_\text{d}\right)$. The calibration factor $\partial \Delta V/\partial x_\text{d}$ is found prior to each experiment by monitoring the change in $\Delta V$ as the photodetector is translated approximately 60 $\upmu$m (the laser spot diameter is roughly 1 mm at the detector) in either direction, mimicking a signal.   

The sum voltage $\Sigma V$ of the photodetector halves is insensitive to the beam deflection and may be used to measure the reflected optical power, where the calibration factor is measured to be $\partial \Sigma V/\partial P_R=9.3$~V/mW via comparison with a power meter.

\section{Mathematical model for radiation pressure experiment}
In Section~\ref{sec:rp}, a radiation pressure drive is used to characterize the torque response of the device. Here we provide the mathematical model for this experiment to motivate the data processing scheme of Fig.~\ref{fig_rp}b.

\subsection{Note on the structural damping model for $\chi_\pm(\omega)$}\label{app_sec:fourier}
Following a common practice for oscillators suspected to be limited by intrinsic material loss~\cite{levin1998internal,numata2004thermal,unterreithmeier2010damping,yu2012control}, we have adopted a structural damping model~\cite{saulson1990thermal} for the mechanical susceptibility $\chi_\pm(\omega)$ (Eq.~\ref{eq:susceptibility}). This model does not correspond to a real-valued system, as $\chi_\pm(\omega)\neq \chi_\pm^*(-\omega)$, particularly noticeable on resonance. This presents a challenge to the analysis of the following sections deriving the time-domain response of the system. However, note that we are primarily concerned with the in-phase response, or the real part of $\chi_\pm(\omega)$, as our driven response measurements in Sections~\ref{sec:rp} and \ref{sec:vib_char} cannot resolve the quadrature component due to its narrow linewidth. 

For the following treatment we assume $H_{\theta \tau}(\omega)=H_{\theta\tau}^*(-\omega)$, such that the angular response $\theta(t)$ to sinusoidal input torques $\tau(t)$ can be explicitly derived in the time domain. For example, given an input torque $\tau(t)=\tau_0 \cos (\omega_\tau t + \phi)$ the angular response comes directly from the inverse Fourier transform
\begin{align*}
    \theta(t) &= \mathcal{F}^{-1} \left\{ H_{\theta \tau}(\omega) \tilde{\tau}(\omega)\right\} \\
    &= \frac{\tau_0}{2} \left( H_{\theta\tau}(\omega_\tau) e^{i\phi}e^{i\omega_\tau t} + H_{\theta\tau}(-\omega_\tau) e^{-i\phi}e^{-i\omega_\tau t} \right)\\
    &\approx H_{\theta \tau}(\omega_\tau) \tau(t)
\end{align*}
where the last line assumes $H_{\theta \tau}(\omega_\tau) \approx H_{\theta \tau}^*(-\omega_\tau)$.

\subsection{Torque and mechanical response}
The laser is focused a distance $l\approx 0.3$ mm from the torsion axis, exerting a radiation pressure torque $\tau(t) = \frac{2l}{c}P_R(t)$ on the test mass due to the reflected laser power $P_R(t)$. The input laser power is modulated sinusoidally by modulating the diode current at a probe frequency $\omega_\text{p}$. Using the same laser beam for detection and excitation presents a challenge to distinguishing the mechanical response from the direct intensity modulation. While a balanced optical lever is immune to laser intensity noise, an imbalance allows the probe tone to be directly measurable in the photodetector difference signal. In theory, the probe intensity modulation would manifest simply as an added constant that could be subtracted. In practice, this added constant is time dependent, due to drift in optical alignment on timescales greater than tens of seconds. To overcome the resulting noise, we inject another modulation tone at $\omega_\text{r}/(2\pi)= 170$ Hz as a reference. 

The total reflected laser power is $P_R(t) = R\left(P_0 + \delta P_\text{p}(t) + \delta P_\text{r}(t)\right)$, where $R\approx 0.2$ is the reflectivity. The angular displacement is then
\begin{equation} \label{eq:angle_response}
    \begin{aligned}
        \theta(t) = \theta_0 + \frac{2lR}{c}\biggl(&H_{\theta\tau}(0) P_0 + H_{\theta\tau}(\omega_\text{p}) \delta P_\text{p}(t) \\
        &+ H_{\theta\tau}(\omega_\text{r}) \delta P_\text{r}(t) \biggr)
    \end{aligned}
\end{equation}
where $\theta_0$ is a constant offset. Strictly speaking, Eq.~\ref{eq:angle_response} applies only to the in-phase response per the simplification described in Section~\ref{app_sec:fourier}.

\subsection{Sum and difference signals}
To first order in $\theta$, the difference signal is
\begin{equation}
    \Delta V(t) = C_\Delta P_R(t) \theta(t)
\end{equation}
where $C_\Delta = \frac{1}{R P_0} \frac{\partial \Delta V}{\partial \theta}$ and the sum signal is  (see Appendix~\ref{sec:OL_calibration}) 
\begin{equation}
    \Sigma V(t) = C_\Sigma P_R(t)
\end{equation}
where $C_\Sigma = \frac{\partial \Sigma V}{\partial P_R}$. As depicted in Fig.~\ref{fig_rp}b, the probe and reference tones are distinguished by bandpass filtering. The filtered signals are
\begin{equation}
    \begin{aligned}
        \Delta V_i(t) &= R C_\Delta \left(\theta_0 + \frac{2lR}{c}P_0\left(H_{\theta\tau}(0)  + H_{\theta\tau}(\omega_i) \right)\right)  \delta P_i(t) \\
        \Sigma V_i(t) &= RC_\Sigma \delta P_i(t)
    \end{aligned}
\end{equation}
where $i\in \left\{ \text{p},\text{r}\right\}$ denotes which tone is targeted. This result assumes clean enough signals so that mixing terms don't have significant components in the pass band. The reference tone frequency of $\omega_\text{r}/2\pi = 170$ Hz was chosen because of the relatively low number of spurious noise peaks within range of the sidebands $102\,\text{Hz}\leq (\omega_\text{r}-\omega_\text{p})/2\pi\leq 135\,\text{Hz}$ and $205\,\text{Hz}\leq (\omega_\text{r}+\omega_\text{p})/2\pi\leq 238\,\text{Hz}$, as the probe tone was swept from 35 Hz to 68 Hz.

\subsection{Lock-in detection} \label{app_sec:rp_LI}
Lock-in detection is performed on the difference signal using the sum signal as a reference over time intervals of $t_\text{LI}$. For example, the in-phase quadratures are
\begin{equation}
    \begin{aligned}
        X_i(t)  &= \frac{\left<\Delta V_i \Sigma V_i\right>_{t_\text{LI}}}{\left<\Sigma V_i^2\right>_{t_\text{LI}}}
    \end{aligned}
\end{equation}
where 
\begin{equation}
    \left<\Delta V_i \Sigma V_i\right>_{t_\text{LI}}\equiv \frac{1}{t_\text{LI}} \int_{t-t_\text{LI}/2}^{t+t_\text{LI}/2} dt' \, \Delta V_i (t') \Sigma V_i (t')
\end{equation}
It can be shown that
\begin{equation}
    X_i = \frac{C_\Delta}{C_\Sigma} \left( \theta_0 + \frac{2lR}{c}P_0 \,\text{Re} \left[H_{\theta\tau}(0)  + H_{\theta\tau}(\omega_i) \right]\right)
\end{equation}
The DC offset terms $\left(\theta_0+\frac{2lR}{c}P_0 H_{\theta\tau}(0)\right)$ allow laser intensity modulation to parasitically couple directly to the difference signal. Due to slow drift in the experiment platform, $\theta_0$ drifts throughout the experiment, resulting in slow drift of $X_i$ between measurements. For this reason the reference tone is included for common mode subtraction
\begin{equation}
    X=\text{cal}\cdot\left(X_\text{p}-X_\text{r}\right) = \text{Re}\left[ H_{\theta\tau}(\omega_\text{p}) - H_{\theta\tau}(\omega_\text{r})\right]  
\end{equation}
where $\text{cal}\equiv \left(\frac{C_\Delta}{C_\Sigma}\frac{2lR}{c}P_0 \right)^{-1}$ is the calibration factor in Fig.~\ref{fig_rp}b, which we estimate from the optical lever calibration described in Appendix~\ref{sec:OL_calibration}. The result acquires a parasitic offset $H_{\theta\tau}(\omega_\text{r})$ that may contain nontrivial contributions from higher-order mechanical modes. Ultimately, we treat the offset as a free parameter $\hat{\beta}$ when fitting the model to the data in Fig.~\ref{fig_rp}c as $X(\omega_\text{p})=\hat{\alpha}\, \text{Re}\left[H_{\theta\tau}(\omega_\text{p})\right]+\hat{\beta}$. The fitted values are $\left(\hat{\alpha},\hat{\beta}\right)=\left(1.0,-5.2\times 10^8 \,{\rm rad/Nm}\right)$ for black and $\left(\hat{\alpha},\hat{\beta}\right)=\left(-1.3,3.8\times 10^8 \,{\rm rad/Nm}\right)$ for orange. Deviation of the scale parameter $\hat{\alpha}$ from 1 or -1 reflects calibration error, and the change in sign between the two measurements reflects the change in sign of the applied radiation pressure torque on opposite sides of the torsion axis.

\bibliography{references.bib}

@article{saulson1990thermal,
  title={Thermal noise in mechanical experiments},
  author={Saulson, Peter R},
  journal={Physical Review D},
  volume={42},
  number={8},
  pages={2437},
  year={1990},
  publisher={APS},
  url={https://journals.aps.org/prd/abstract/10.1103/PhysRevD.42.2437}
}

@inproceedings{pratt2023intersection,
  title={The intersection of noise, amplitude, and nonlinearity in a high-Q micromechanical torsion pendulum},
  author={Pratt, Jon R and Schlamminger, Stephan and Agrawal, Aman R and Condos, Charles A and Pluchar, Christian M and Wilson, Dalziel J},
  booktitle={International Conference on Nonlinear Dynamics and Applications},
  pages={3--14},
  year={2023},
  organization={Springer},
  url= {https://link.springer.com/chapter/10.1007/978-3-031-50635-2_1}
}

@article{westphal2021measurement,
  title={Measurement of gravitational coupling between millimetre-sized masses},
  author={Westphal, Tobias and Hepach, Hans and Pfaff, Jeremias and Aspelmeyer, Markus},
  journal={Nature},
  volume={591},
  number={7849},
  pages={225--228},
  year={2021},
  publisher={Nature Publishing Group UK London},
  url={https://www.nature.com/articles/s41586-021-03250-7}
}

@article{bsaibes2025lithographically,
  title={Lithographically Defined Si$_3$N$_4$ Torsional Pendulum},
  author={Bsaibes, T and Condos, C and Manley, J and Pratt, J and Wilson, D and Taylor, J},
  journal={arXiv preprint arXiv:2512.13435},
  year={2025},
  url={https://arxiv.org/abs/2512.13435}
}

@article{condos2025ultralow,
  title={Ultralow loss torsion micropendula for chipscale gravimetry},
  author={Condos, CA and Pratt, JR and Manley, J and Agrawal, AR and Schlamminger, S and Pluchar, CM and Wilson, DJ},
  journal={Physical Review Letters},
  volume={134},
  number={25},
  pages={253602},
  year={2025},
  publisher={APS},
  url={https://journals.aps.org/prl/abstract/10.1103/nmx5-hygh}
}

@article{sokhi2024all,
  title={All optical zepto-Newton-meter nanoscale silk sensor},
  author={Sokhi, Shivali and Singh, Kamal P},
  journal={Physical Review Letters},
  volume={133},
  number={8},
  pages={083801},
  year={2024},
  publisher={APS},
  url={https://journals.aps.org/prl/abstract/10.1103/PhysRevLett.133.083801}
}

@article{guan2026all,
  title={All-Optically Operated Atto-Newton Force Sensing with a Centimeter-Milligram-Scale Torsion Pendulum},
  author={Guan, Sheng-Guo and Cheng, Yan-Bei and Sun, Jing and Duan, Zheng-Lu and Le, Jian-Xin},
  journal={Physical Review Letters},
  volume={136},
  number={6},
  pages={063603},
  year={2026},
  publisher={APS},
  url={https://journals.aps.org/prl/abstract/10.1103/2qrz-5b94}
}

@article{prasad202219,
  title={A 19 day earth tide measurement with a MEMS gravimeter},
  author={Prasad, Abhinav and Middlemiss, Richard P and Noack, Andreas and Anastasiou, Kristian and Bramsiepe, Steven G and Toland, Karl and Utting, Phoebe R and Paul, Douglas J and Hammond, Giles D},
  journal={Scientific reports},
  volume={12},
  number={1},
  pages={13091},
  year={2022},
  publisher={Nature Publishing Group UK London},
  url={https://www.nature.com/articles/s41598-022-16881-1}
}

@article{liu2001modes,
  title={On the modes and loss mechanisms of a high Q mechanical oscillator},
  author={Liu, Xiao and Morse, SF and Vignola, JF and Photiadis, DM and Sarkissian, A and Marcus, MH and Houston, BH},
  journal={Applied Physics Letters},
  volume={78},
  number={10},
  pages={1346--1348},
  year={2001},
  publisher={American Institute of Physics},
  url={https://pubs.aip.org/aip/apl/article/78/10/1346/113042/On-the-modes-and-loss-mechanisms-of-a-high-Q}
}

@article{cong2021chip,
  title={On-chip torsion balances with femtonewton force resolution at room temperature enabled by carbon nanotube and graphene},
  author={Cong, Lin and Yuan, Zi and Bai, Zaiqiao and Wang, Xinhe and Zhao, Wei and Gao, Xinyu and Hu, Xiaopeng and Liu, Peng and Guo, Wanlin and Li, Qunqing and others},
  journal={Science Advances},
  volume={7},
  number={12},
  pages={eabd2358},
  year={2021},
  publisher={American Association for the Advancement of Science},
  url={https://www.science.org/doi/full/10.1126/sciadv.abd2358}
}

@article{agafonova2026one,
  title={One-milligram torsional pendulum toward experiments at the quantum-gravity interface},
  author={Agafonova, Sofia and Rossell{\'o}, Pere and Mekonnen, Manuel and Hosten, Onur},
  journal={Communications Physics},
  year={2026},
  publisher={Nature Publishing Group UK London},
  url={https://www.nature.com/articles/s42005-026-02514-w}
}

@inproceedings{chandak2026nano,
  title={Nano-g optomechanical accelerometry with a chip-scale torsion pendulum},
  author={Chandak, Namit and Condos, Charles A and Manley, Jack and Pratt, Jon R and Wilson, Dalziel J},
  booktitle={Quantum Sensing, Imaging, and Precision Metrology IV},
  volume={13920},
  pages={31--37},
  year={2026},
  organization={SPIE},
  url={https://www.spiedigitallibrary.org/conference-proceedings-of-spie/13920/1392006/Nano-g-optomechanical-accelerometry-with-a-chip-scale-torsion-pendulum/10.1117/12.3091430.short}
}

@article{komori2020attonewton,
  title={Attonewton-meter torque sensing with a macroscopic optomechanical torsion pendulum},
  author={Komori, Kentaro and Enomoto, Yutaro and Ooi, Ching Pin and Miyazaki, Yuki and Matsumoto, Nobuyuki and Sudhir, Vivishek and Michimura, Yuta and Ando, Masaki},
  journal={Physical Review A},
  volume={101},
  number={1},
  pages={011802},
  year={2020},
  publisher={APS},
  url={https://journals.aps.org/pra/abstract/10.1103/PhysRevA.101.011802}
}

@article{manley2024microscale,
  title={Microscale torsion resonators for short-range gravity experiments},
  author={Manley, J and Condos, CA and Schlamminger, S and Pratt, JR and Wilson, DJ and Terrano, WA},
  journal={Physical Review D},
  volume={110},
  number={12},
  pages={122005},
  year={2024},
  publisher={APS},
  url={https://journals.aps.org/prd/abstract/10.1103/PhysRevD.110.122005}
}

@article{sundrum2004fat,
  title={Fat gravitons, the cosmological constant and submillimeter tests},
  author={Sundrum, Raman},
  journal={Physical Review D},
  volume={69},
  number={4},
  pages={044014},
  year={2004},
  publisher={APS},
  url={https://journals.aps.org/prd/abstract/10.1103/PhysRevD.69.044014}
}

@article{sadeghi2020frequency,
  title={Frequency fluctuations in nanomechanical silicon nitride string resonators},
  author={Sadeghi, Pedram and Demir, Alper and Villanueva, Luis Guillermo and K{\"a}hler, Hendrik and Schmid, Silvan},
  journal={Phys. Rev. B},
  volume={102},
  number={21},
  pages={214106},
  year={2020},
  publisher={APS},
url={https://journals.aps.org/prb/abstract/10.1103/PhysRevB.102.214106}
}

@article{upadhye2012chameleon,
  title = {Quantum Stability of Chameleon Field Theories},
  author = {Upadhye, Amol and Hu, Wayne and Khoury, Justin},
  journal = {Phys. Rev. Lett.},
  volume = {109},
  issue = {4},
  pages = {041301},
  numpages = {5},
  year = {2012},
  month = {Jul},
  publisher = {American Physical Society},
  doi = {10.1103/PhysRevLett.109.041301},
  url = {https://link.aps.org/doi/10.1103/PhysRevLett.109.041301}
}

@article{montero2023dark,
  title={The dark dimension and the Swampland},
  author={Montero, Miguel and Vafa, Cumrun and Valenzuela, Irene},
  journal={Journal of High Energy Physics},
  volume={2023},
  number={2},
  pages={1--18},
  year={2023},
  publisher={Springer},
  url={https://link.springer.com/article/10.1007/JHEP02(2023)022}
}

@article{pratt2023nanoscale,
  title={Nanoscale torsional dissipation dilution for quantum experiments and precision measurement},
  author={Pratt, Jon R and Agrawal, Aman R and Condos, Charles A and Pluchar, Christian M and Schlamminger, Stephan and Wilson, Dalziel J},
  journal={Physical Review X},
  volume={13},
  number={1},
  pages={011018},
  year={2023},
  publisher={APS},
  url={https://journals.aps.org/prx/abstract/10.1103/PhysRevX.13.011018}
}

@article{yu2012control,
  title={Control of material damping in high-Q membrane microresonators},
  author={Yu, P-L and Purdy, TP and Regal, CA},
  journal={Physical review letters},
  volume={108},
  number={8},
  pages={083603},
  year={2012},
  publisher={APS},
  url={https://journals.aps.org/prl/abstract/10.1103/PhysRevLett.108.083603}
}

@article{unterreithmeier2010damping,
  title={Damping of nanomechanical resonators},
  author={Unterreithmeier, Quirin P and Faust, Thomas and Kotthaus, J{\"o}rg P},
  journal={Physical review letters},
  volume={105},
  number={2},
  pages={027205},
  year={2010},
  publisher={APS},
  url={https://journals.aps.org/prl/abstract/10.1103/PhysRevLett.105.027205}
}

@article{numata2004thermal,
  title={Thermal-noise limit in the frequency stabilization of lasers with rigid cavities},
  author={Numata, Kenji and Kemery, Amy and Camp, Jordan},
  journal={Physical review letters},
  volume={93},
  number={25},
  pages={250602},
  year={2004},
  publisher={APS},
  url={https://journals.aps.org/prl/abstract/10.1103/PhysRevLett.93.250602}
}

@inproceedings{des2025gravity,
  title={Gravity Measurement with a Quantum Inertial Sensor for Mobile Applications},
  author={Des Cognets, Cyrille and Lenogue, Guillaume and de Castanet, Quentin d'Armagnac and Jarlaud, Vincent and Menoret, Vincent and Battelier, Baptiste},
  booktitle={2025 IEEE International Symposium on Inertial Sensors and Systems (INERTIAL)},
  pages={1--4},
  year={2025},
  organization={IEEE},
  url={https://ieeexplore.ieee.org/abstract/document/11037179}
}

@article{levin1998internal,
  title={Internal thermal noise in the LIGO test masses: A direct approach},
  author={Levin, Yu},
  journal={Physical Review D},
  volume={57},
  number={2},
  pages={659},
  year={1998},
  publisher={APS},
  url={https://journals.aps.org/prd/abstract/10.1103/PhysRevD.57.659}
}

@article{zhou2025stability,
  title={Stability of Shipborne Quantum Gravimeter in Long-term Continuous Marine Gravity Measurement},
  author={Zhou, Yin and Zhang, Can and Chen, Qianlong and Wang, Zhenghao and Wang, Kainan and Ma, Zhixiang and Yang, Yi and Li, Rui and Wu, Bin and Wang, Xiaolong and others},
  journal={IEEE Sensors Journal},
  year={2025},
  publisher={IEEE},
  url={https://ieeexplore.ieee.org/abstract/document/11165789}
}

@article{middlemiss2016measurement,
  title={Measurement of the Earth tides with a MEMS gravimeter},
  author={Middlemiss, RP and Samarelli, Antonio and Paul, DJ and Hough, James and Rowan, Sheila and Hammond, GD},
  journal={Nature},
  volume={531},
  number={7596},
  pages={614--617},
  year={2016},
  publisher={Nature Publishing Group UK London},
  url={https://www.nature.com/articles/nature17397}
}

@article{tang2019high,
  title={A high-sensitivity MEMS gravimeter with a large dynamic range},
  author={Tang, Shihao and Liu, Huafeng and Yan, Shitao and Xu, Xiaochao and Wu, Wenjie and Fan, Ji and Liu, Jinquan and Hu, Chenyuan and Tu, Liangcheng},
  journal={Microsystems \& nanoengineering},
  volume={5},
  number={1},
  pages={45},
  year={2019},
  publisher={Nature Publishing Group UK London},
  url={https://www.nature.com/articles/s41378-019-0089-7}
}

@article{gao2026force,
  title={Force balanced chip scale gravimeter achieving record low self noise of 0.1 micro-Gal/rtHz},
  author={Gao, Le and Wu, WenJie and Li, FangZheng and Cai, Bingyang and Xie, RunHan and Zhang, Zhong and Yang, LuJia and Zhang, Jian and Peng, MaoJun and Wang, Yuan and others},
  journal={Microsystems \& Nanoengineering},
  volume={12},
  number={1},
  pages={8},
  year={2026},
  publisher={Nature Publishing Group UK London},
  url={https://www.nature.com/articles/s41378-025-01039-6}
}

@article{zhou2021broadband,
  title={Broadband thermomechanically limited sensing with an optomechanical accelerometer},
  author={Zhou, Feng and Bao, Yiliang and Madugani, Ramgopal and Long, David A and Gorman, Jason J and LeBrun, Thomas W},
  journal={Optica},
  volume={8},
  number={3},
  pages={350--356},
  year={2021},
  publisher={Optical Society of America},
  url={https://opg.optica.org/optica/fulltext.cfm?uri=optica-8-3-350}
}

@article{bawden2025precision,
  title={Precision optomechanical accelerometer via hybrid test-mass integration},
  author={Bawden, Nathaniel and Carey, Benjamin J and Yeo, Poh-Meng and Arora, Nishta and Sementilli, Leo and Valenzuela, Victor M and Romero, Erick and Harris, Glen I and Wegener, Margaret and Bowen, Warwick P},
  journal={Physical Review Applied},
  volume={24},
  number={6},
  pages={064008},
  year={2025},
  publisher={APS},
  url={https://journals.aps.org/prapplied/abstract/10.1103/knpw-1mdj}
}

@article{guzman2014high,
  title={High sensitivity optomechanical reference accelerometer over 10 kHz},
  author={Guzm{\'a}n Cervantes, Felipe and Kumanchik, Lee and Pratt, Jon and Taylor, Jacob M},
  journal={Applied Physics Letters},
  volume={104},
  number={22},
  year={2014},
  publisher={AIP Publishing},
  url={https://pubs.aip.org/aip/apl/article/104/22/221111/26557}
}

@article{li20222,
  title={2 ng/rtHz-resolution optomechanical accelerometer employing a three-dimensional MEMS interferometer},
  author={Li, Cheng and Yang, Bo and Zheng, Xiang and Guo, Xin and Sun, Zhenyu and Zhou, Luqiang and Huang, Xin},
  journal={Optics Letters},
  volume={47},
  number={7},
  pages={1883--1886},
  year={2022},
  publisher={Optica Publishing Group},
  url={https://opg.optica.org/ol/fulltext.cfm?uri=ol-47-7-1883}
}

@article{weidong2025low,
  title={A low-noise MEMS gravimeter capable of attaining stable low resonant frequency at various tilt angles through adjusting curved beams’ width},
  author={Weidong, Wang and Minghong, Qi and Shilong, Jin and Taotao, Ding and Zhengqian, Zhao and Zhi, Ding and Ye, Zhu and Baoyin, Hou and Lufeng, Che},
  journal={IEEE Transactions on Electron Devices},
  volume={72},
  number={3},
  pages={1368--1376},
  year={2025},
  publisher={IEEE},
  url={https://ieeexplore.ieee.org/abstract/document/10858444}
}

@article{jiao2024optomechanical,
  title={An optomechanical MEMS geophone with a 2.5 ng/Hz1/2 noise floor for oil/gas exploration},
  author={Jiao, Shimin and Qu, Ziqiang and Ma, Xujin and Ouyang, Hao and Xiong, Wen and Zhang, Shaolin and Wang, Qiu and Liu, Huafeng},
  journal={Microsystems \& Nanoengineering},
  volume={10},
  number={1},
  pages={176},
  year={2024},
  publisher={Nature Publishing Group UK London},
  url={https://www.nature.com/articles/s41378-024-00802-5}
}

@article{wang2025measurement,
  title={Measurement of Earth tides with a MEMS gravimeter via releasing the axial force of cross-configured geometric anti-spring},
  author={Wang, Weidong and Ding, Taotao and Chen, Yiming and Song, Mengxuan and Li, Rui and Leng, Yingchun and Qi, Minghong and Huang, Pu and Li, Ying and Che, Lufeng},
  journal={Journal of Micromechanics and Microengineering},
  volume={35},
  number={9},
  pages={095012},
  year={2025},
  publisher={IOP Publishing},
  url={https://iopscience.iop.org/article/10.1088/1361-6439/ae0488/meta}
}

@article{mustafazade2020vibrating,
  title={A vibrating beam MEMS accelerometer for gravity and seismic measurements},
  author={Mustafazade, Arif and Pandit, Milind and Zhao, Chun and Sobreviela, Guillermo and Du, Zhijun and Steinmann, Philipp and Zou, Xudong and Howe, Roger T and Seshia, Ashwin A},
  journal={Scientific reports},
  volume={10},
  number={1},
  pages={10415},
  year={2020},
  publisher={Nature Publishing Group UK London},
  url={https://www.nature.com/articles/s41598-020-67046-x}
}

@article{smullin2005constraints,
  title={Constraints on Yukawa-type deviations from Newtonian gravity at 20 microns},
  author={Smullin, SJ and Geraci, AA and Weld, DM and Chiaverini, J and Holmes, S and Kapitulnik, A},
  journal={Physical Review D},
  volume={72},
  number={12},
  pages={122001},
  year={2005},
  publisher={APS},
  url={https://journals.aps.org/prd/abstract/10.1103/PhysRevD.72.122001}
}

@article{dey2026optomechanical,
  title={Optomechanical accelerometer search for ultralight dark matter},
  author={Dey Chowdhury, M and Manley, JP and Condos, CA and Agrawal, AR and Wilson, DJ},
  journal={Physical Review D},
  volume={113},
  number={12},
  pages={L121303},
  year={2026},
  publisher={APS},
  url={https://journals.aps.org/prd/abstract/10.1103/16yd-qj5v}
}

@article{fischbach2001new,
  title={New constraints on ultrashort-ranged {Y}ukawa interactions from atomic force microscopy},
  author={Fischbach, Ephraim and Krause, DE and Mostepanenko, VM and Novello, M},
  journal={Physical Review D},
  volume={64},
  number={7},
  pages={075010},
  year={2001},
  publisher={APS},
  url={https://journals.aps.org/prd/abstract/10.1103/PhysRevD.64.075010}
}

@article{adelberger2007implications,
  title = {Particle-Physics Implications of a Recent Test of the Gravitational Inverse-Square Law},
  author = {Adelberger, E. G. and Heckel, B. R. and Hoedl, S. and Hoyle, C. D. and Kapner, D. J. and Upadhye, A.},
  journal = {Phys. Rev. Lett.},
  volume = {98},
  issue = {13},
  pages = {131104},
  numpages = {4},
  year = {2007},
  month = {Mar},
  publisher = {American Physical Society},
  doi = {10.1103/PhysRevLett.98.131104},
  url = {https://link.aps.org/doi/10.1103/PhysRevLett.98.131104}
}

@article{adelberger2009torsion,
  title={Torsion balance experiments: A low-energy frontier of particle physics},
  author={Adelberger, Eric G and Gundlach, JH and Heckel, BR and Hoedl, S and Schlamminger, S},
  journal={Progress in Particle and Nuclear Physics},
  volume={62},
  number={1},
  pages={102--134},
  year={2009},
  publisher={Elsevier},
  url={https://www.sciencedirect.com/science/article/pii/S0146641008000720}
}

@article{geraci2008improved,
  title={Improved constraints on non-Newtonian forces at 10 microns},
  author={Geraci, Andrew A and Smullin, Sylvia J and Weld, David M and Chiaverini, John and Kapitulnik, Aharon},
  journal={Physical Review D},
  volume={78},
  number={2},
  pages={022002},
  year={2008},
  publisher={APS},
  url={https://journals.aps.org/prd/abstract/10.1103/PhysRevD.78.022002}
}

@article{arkani1999phenomenology,
  title={Phenomenology, astrophysics, and cosmology of theories with submillimeter dimensions and TeV scale quantum gravity},
  author={Arkani-Hamed, Nima and Dimopoulos, Savas and Dvali, Gia},
  journal={Physical Review D},
  volume={59},
  number={8},
  pages={086004},
  year={1999},
  publisher={APS},
  url={https://journals.aps.org/prd/abstract/10.1103/PhysRevD.59.086004}
}

@article{catano2020high,
  title={High-q milligram-scale monolithic pendulum for quantum-limited gravity measurements},
  author={Cata{\~n}o-Lopez, Seth B and Santiago-Condori, Jordy G and Edamatsu, Keiichi and Matsumoto, Nobuyuki},
  journal={Physical review letters},
  volume={124},
  number={22},
  pages={221102},
  year={2020},
  publisher={APS},
  url={https://journals.aps.org/prl/abstract/10.1103/PhysRevLett.124.221102}
}

@article{manley2026nanofabricated,
  title={Nanofabricated torsion-pendulum suspensions for tabletop gravity experiments},
  author={Manley, J and Condos, CA and Fegley, Z and Premawardhana, G and Bsaibes, T and Taylor, JM and Wilson, DJ and Pratt, JR},
  journal={Physical Review Applied},
  volume={25},
  number={5},
  pages={054033},
  year={2026},
  publisher={APS},
  url={https://journals.aps.org/prapplied/abstract/10.1103/mnrd-3bm2}
}

@article{matichard2015seismic,
  title={Seismic isolation of Advanced LIGO: Review of strategy, instrumentation and performance},
  author={Matichard, Fabrice and Lantz, B and Mittleman, Richard and Mason, Kenneth and Kissel, J and Abbott, B and Biscans, S and McIver, J and Abbott, R and Abbott, S and others},
  journal={Classical and Quantum Gravity},
  volume={32},
  number={18},
  pages={185003},
  year={2015},
  publisher={IOP Publishing},
  url={https://iopscience.iop.org/article/10.1088/0264-9381/32/18/185003/meta}
}

@article{farah2014underground,
  title={Underground operation at best sensitivity of the mobile LNE-SYRTE cold atom gravimeter},
  author={Farah, Tristan and Guerlin, Christine and Landragin, Arnaud and Bouyer, Ph and Gaffet, St{\'e}phane and Pereira Dos Santos, Franck and Merlet, S{\'e}bastien},
  journal={Gyroscopy and Navigation},
  volume={5},
  number={4},
  pages={266--274},
  year={2014},
  publisher={Springer},
  url={https://link.springer.com/article/10.1134/S2075108714040051}
}

@article{wu2024construction,
  title={Construction of a test field for relative gravimeters in a cave with a cold atom gravimeter},
  author={Wu, Bin and Li, Dianrong and Zhou, Yin and Zhu, Dong and Zhao, Yingpeng and Qiao, Zhongkun and Cheng, Bing and Niu, Jingyu and Guo, Xiaochun and Wang, Xiaolong and others},
  journal={IEEE Sensors Journal},
  volume={24},
  number={7},
  pages={9536--9544},
  year={2024},
  publisher={IEEE},
  url={https://ieeexplore.ieee.org/document/10439016}
}

@article{venugopalan2026optomechanical,
  title={Optomechanical vector sensing of new forces at 6 micron separation},
  author={Venugopalan, Gautam and Hardy, Clarke A and Kohn, Kenneth and Zhu, Yuqi and Blakemore, Charles P and Fieguth, Alexander and Huang, Jacqueline and Jia, Chengjie and Liu, Meimei and Magrini, Lorenzo and others},
  journal={Scientific Reports},
  year={2026},
  publisher={Nature Publishing Group UK London},
  url={https://www.nature.com/articles/s41598-026-35656-6}
}

@article{antoniadis2003brane,
  title={Brane to bulk supersymmetry breaking and radion force at micron distances},
  author={Antoniadis, Ignatios and Benakli, Karim and Laugier, Alexander and Maillard, Tristan},
  journal={Nuclear Physics B},
  volume={662},
  number={1-2},
  pages={40--62},
  year={2003},
  publisher={Elsevier},
  url={https://www.sciencedirect.com/science/article/pii/S0550321303002554}
}

@article{kaplan2000couplings,
  title={Couplings of a light dilaton and violations of the equivalence principle},
  author={Kaplan, David B and Wise, Mark B},
  journal={Journal of High Energy Physics},
  volume={2000},
  number={08},
  pages={037},
  year={2000},
  publisher={IOP Publishing},
  url={https://iopscience.iop.org/article/10.1088/1126-6708/2000/08/037/meta}
}

@article{chen2016stronger,
  title={Stronger limits on hypothetical Yukawa interactions in the 30--8000 nm range},
  author={Chen, Y-J and Tham, Weng K and Krause, DE and L{\'o}pez, D and Fischbach, Ephraim and Decca, Ricardo S},
  journal={Physical Review Letters},
  volume={116},
  number={22},
  pages={221102},
  year={2016},
  publisher={APS},
  url={https://journals.aps.org/prl/abstract/10.1103/PhysRevLett.116.221102}
}

@article{adelberger2003tests,
  title={Tests of the gravitational inverse-square law},
  author={Adelberger, EG and Heckel, BR and Nelson, AE},
  journal={Annual Review of Nuclear and Particle Science},
  volume={53},
  pages={77},
  year={2003},
  publisher={Annual Reviews, Inc.},
  url = {https://www.annualreviews.org/content/journals/10.1146/annurev.nucl.53.041002.110503}
}

@article{Kapner2007,
  title = {Tests of the Gravitational Inverse-Square Law below the Dark-Energy Length Scale},
  author = {Kapner, D. J. and Cook, T. S. and Adelberger, E. G. and Gundlach, J. H. and Heckel, B. R. and Hoyle, C. D. and Swanson, H. E.},
  journal = {Phys. Rev. Lett.},
  volume = {98},
  issue = {2},
  pages = {021101},
  numpages = {4},
  year = {2007},
  month = {Jan},
  publisher = {American Physical Society},
  doi = {10.1103/PhysRevLett.98.021101},
  url = {https://link.aps.org/doi/10.1103/PhysRevLett.98.021101}
}

@article{blakemore2021search,
  title={Search for non-Newtonian interactions at micrometer scale with a levitated test mass},
  author={Blakemore, Charles P and Fieguth, Alexander and Kawasaki, Akio and Priel, Nadav and Martin, Denzal and Rider, Alexander D and Wang, Qidong and Gratta, Giorgio},
  journal={Physical Review D},
  volume={104},
  number={6},
  pages={L061101},
  year={2021},
  publisher={APS},
  url={https://journals.aps.org/prd/abstract/10.1103/PhysRevD.104.L061101}
}

@article{lee2020new,
  title={New test of the gravitational 1/r$^2$ law at separations down to 52$\,\upmu$m},
  author={Lee, JG and Adelberger, EG and Cook, TS and Fleischer, SM and Heckel, BR},
  journal={Physical Review Letters},
  volume={124},
  number={10},
  pages={101101},
  year={2020},
  publisher={APS},
  url={https://journals.aps.org/prl/abstract/10.1103/PhysRevLett.124.101101}
}

@article{tan2020improvement,
  title={Improvement for testing the gravitational inverse-square law at the submillimeter range},
  author={Tan, Wen-Hai and Du, An-Bin and Dong, Wen-Can and Yang, Shan-Qing and Shao, Cheng-Gang and Guan, Sheng-Guo and Wang, Qing-Lan and Zhan, Bi-Fu and Luo, Peng-Shun and Tu, Liang-Cheng and others},
  journal={Physical Review Letters},
  volume={124},
  number={5},
  pages={051301},
  year={2020},
  publisher={APS},
  url={https://journals.aps.org/prl/abstract/10.1103/PhysRevLett.124.051301}
}

@article{long2003upper,
  title={Upper limits to submillimetre-range forces from extra space-time dimensions},
  author={Long, Joshua C and Chan, Hilton W and Churnside, Allison B and Gulbis, Eric A and Varney, Michael and Price, John C},
  journal={Nature},
  volume={421},
  number={6926},
  pages={922--925},
  year={2003},
  publisher={Nature Publishing Group},
  url={https://www.nature.com/articles/nature01432}
}
\end{document}